\documentclass[
    ,final            % use final for the camera ready runs
  ]
  {aipproc}

\layoutstyle{6x9}

\usepackage{natbib}

\def\ls{LS~5039}
\def\lsi{LSI~$+$61$^{\rm{o}}$303}
\def\psrb{PSR~B1259$-$63}
\def\j0632{HESS~J0632$+$057}

\begin{document}

\title{What can {\em Simbol-X} do for gamma-ray binaries?}


\author{B. Cerutti}{
  address={Laboratoire d'Astrophysique de Grenoble, UMR 5571 CNRS, Universit\'e
Joseph Fourier, BP 53, 38041 Grenoble, France}
}

\author{G. Dubus}{
  address={Laboratoire d'Astrophysique de Grenoble, UMR 5571 CNRS, Universit\'e
Joseph Fourier, BP 53, 38041 Grenoble, France}
}

\author{G. Henri}{
  address={Laboratoire d'Astrophysique de Grenoble, UMR 5571 CNRS, Universit\'e
Joseph Fourier, BP 53, 38041 Grenoble, France}
}
\author{A. B. Hill}{
  address={Laboratoire d'Astrophysique de Grenoble, UMR 5571 CNRS, Universit\'e
Joseph Fourier, BP 53, 38041 Grenoble, France}
}
\author{A. Szostek}{
  address={Laboratoire d'Astrophysique de Grenoble, UMR 5571 CNRS, Universit\'e
Joseph Fourier, BP 53, 38041 Grenoble, France}
}

\begin{abstract}
Gamma-ray binaries have been uncovered as a new class of Galactic objects in the very high energy sky (> 100 GeV). The three systems known today have hard X-ray spectra (photon index $\sim$1.5), extended radio emission and a high luminosity in gamma-rays. Recent monitoring campaigns of \lsi\ in X-rays have confirmed variability in these systems and revealed a spectral hardening with increasing flux. In a generic one-zone leptonic model, the cooling of relativistic electrons accounts for the main spectral and temporal features observed at high energy. Persistent hard X-ray emission is expected to extend well beyond 10 keV. We explain how {\em Simbol-X} will constrain the existing models in connection with {\em Fermi Space Telescope} measurements. Because of its unprecedented sensitivity in hard X-rays, {\em Simbol-X} will also play a role in the discovery of new gamma-ray binaries, giving new insights into the evolution of compact binaries.
\end{abstract}
%In the pulsar model, the collision between the winds of the massive star and a young pulsar accounts for the main spectral and temporal features
\maketitle

%%%%%%%%%%%%%%%%%%%%%%%%%%%%%%%%%%%%%%%%%%%%
%% MAINMATTER
%%%%%%%%%%%%%%%%%%%%%%%%%%%%%%%%%%%%%%%%%%%%

\section{Introduction}

Initially classified as standard high-mass X-ray binaries, gamma-ray binaries were soon set apart because of their low X-ray luminosity ($\rm{L_X}\sim 10^{33}$-$10^{34}$ erg/s) and their resolved radio counter part. The new generation of Cherenkov telescopes has unveiled the real nature of these systems and established a new class of Galactic objects. Gamma-ray binaries emit most of their radiative output in the gamma-ray energy band from MeV up to about 10 TeV. They appear as point-like sources and exhibit an orbital modulated flux in the very high energy sky (VHE >100 GeV). Three systems are now solidly identified as gamma-ray binaries: \psrb\  and \ls, both discovered by HESS in the Galactic Plane \citep{2005A&A...442....1A,2005Sci...309..746A} and \lsi\ by MAGIC \citep{2006Sci...312.1771A}, later on confirmed by VERITAS \citep{2007arXiv0709.3661M}. Serendipitously detected by HESS \citep{2007A&A...469L...1A}, \j0632\ is possibly the fourth system discovered so far \citep{2008arXiv0809.0584H} but more investigations are necessary to confirm its nature. \psrb\ is comprised of a young rotation-powered 48 ms pulsar and a massive Be-star, but the nature of the compact object in the other systems remains unknown \footnote{A recent soft gamma-ray repeater/anomalous X-ray pulsar like burst was observed by {\em SWIFT} (BAT) in the direction of \lsi, possibly betraying the activity of a magnetar in this system \citep{2008ATel.1715....1D}.}.\\

Particles in gamma-ray binaries are accelerated with high efficiency to several TeV energies, but the underlying physical mechanisms are still poorly understood. These systems are ideal objects for modeling the non-thermal radiation. The soft photon density is set by the massive star black body spectrum and the geometry of the system given by the orbital parameters. Observations in hard X-rays (20-100 keV) can constrain the existing models but suffer from low sensitivity of current telescopes. With its unprecedented sensitivity {\em Simbol-X} will be able to better understand the physics at work in these systems, giving stringent constraints on models. The study of these systems offers the opportunity to explore a new class of Galactic object. We also explain here how {\em Simbol-X} will identify new candidates and possibly serendipitously discover new systems.

\section{{\em Simbol-X} can constrain existing models}

\begin{figure}[h]
   \centering
   \includegraphics*[width=14cm]{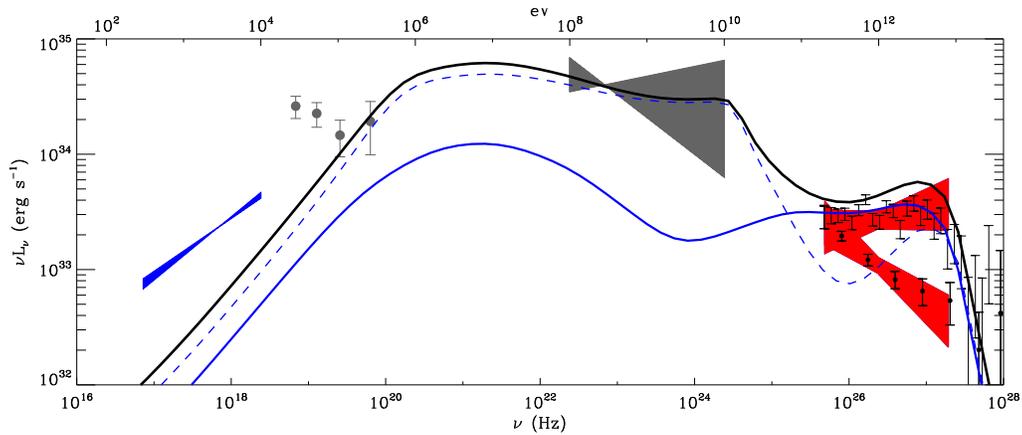}
      \caption{Non-thermal radiation from \ls\ expected in a one-zone leptonic model. Orbital modulation is expected from X-rays up to VHE. Orbit (solid black line), superior conjunction (SUPC) (blue dashed line) and inferior conjunction (INFC) (blue solid line) averaged spectra. See \citep{2008A&A...477..691D} for more details.}
   \label{b_cerutti_f1}
\end{figure}

Independent of the nature of the compact object, the main spectral features exhibited by gamma-ray binaries at high energy can be explained by a radiating population of relativistic electrons. In a generic one zone leptonic model, particles are injected in a region bathed in an ambient magnetic field and by soft photons from the massive star. The non-thermal radiation is produced by synchrotron radiation and inverse Compton scattering of the high energy electrons onto the soft stellar photons. Taking into account the anisotropic effects due to the relative position of the compact object and the massive star to the observer, spectra averaged along the orbit can be computed (Fig. \ref{b_cerutti_f1}).

In this model, a break energy is expected at the transition between
the hard X-rays and the high energy band (HE, GeV domain). This break
is related to the synchrotron radiation of the electrons and depends
on the magnetic field strength as %$\epsilon_{sync}\propto 1/B$.
\begin{equation}
\epsilon_{sync}=750\left(\frac{T_{\star}}{4\times10^{4}~\rm{K}}\right)^2\left(\frac{R_{\star}}{10~R_{\odot}}\right)^2\left(\frac{d}{0.1~\rm{AU}}\right)^{-2}\left(\frac{B}{1~\rm{G}}\right)^{-1}~\rm{keV},
\end{equation}
where d is the orbital separation. Hence, {\em Simbol-X} observations in the hard X-ray band in connection with the measurements of the {\em Fermi Space Telescope} at HE will constrain the magnetic field.\\

Low variability correlated with the orbital period is expected. X-ray observations of gamma-ray binaries have not revealed a clear orbital modulation until recently \citep{2008arXiv0812.0766H,2008arXiv0812.3358T} for \ls\ (see the contribution by M. Chernyakova in this volume for \psrb). Some binaries present more complicated patterns in their lightcurves. Long term {\em RXTE} monitoring of \lsi\ has revealed flaring episodes with no orbital phase correlations \citep{2008arXiv0809.4254S}. Spectral hardening with increasing flux have also been reported in the 2-10 keV band. Three flares of an hour long were clearly observed where the flux is one order of magnitude higher than the quiescence state. The doubling time is about 2 s, constraining the emitting size region to $\sim$ $6\times10^{10}$ cm, consistent with the expected size of the pulsar wind zone \citep{2006A&A...456..801D}. The sub-orbital variability might be due to the interaction with a clump from the massive star \citep{2008arXiv0802.1174Z}. {\em Simbol-X} will certainly help to interpret this puzzling behaviour.

\section{{\em Simbol-X} can identify new candidates}

\subsection{Candidates for follow-up}

A straightforward way to find new candidates in the hard X-ray sky is to look at the unidentified sources of the {\em INTEGRAL} (IBIS/ISGRI) survey catalog. About 25\% of the sources detected are unidentified \citep{2007ApJS..170..175B} and this fraction will probably remain unchanged by the end of the mission in 2012. The known gamma-ray binaires exhibit a rather hard (photon index $\sim 1.5$) steady spectrum with no high energy cut-off in hard X-rays observed so far. However, these criteria are not sufficiently discriminative to identify good candidates. Multi-catalog cross correlations are necessary to reduce the number of potential targets for {\em Simbol-X}. The first step is to look for a massive star coincident with the X-ray source and if possible find the orbital period of the system. Searching for an (extended) radio source increases the probability to find a candidate. Cross correlations with the {\em Fermi} and/or the Cherenkov telescopes observations in the Galactic Plane will definitively identify good candidates for follow-up by {\em Simbol-X}.

\subsection{Serendipitous detections}

Even if {\em Simbol-X} was not designed for a sky survey, the telescope will be pointed several times in the Galactic Plane. Here, we would like to give a rough estimate of potential serendipitous discoveries of new gamma-ray binaires with {\em Simbol-X}. The HESS survey of the Galactic Plane \citep{2006ApJ...636..777A} has discovered two, perhaps three systems with a limiting flux of about $\rm{F_{VHE}}$ $>10^{-12}$ erg $\rm{cm}^{-2}$ $\rm{s}^{-1}$ (2\% Crab), over an area of about 300 deg$^2$. Assuming 500 pointings of 20 ks in the Galactic Plane, {\em Simbol-X} will cover about 20 deg$^{2}$ with a limiting flux $\rm{F_{X}}$(20-40 keV) $>10^{-13}$ erg $\rm{cm}^{-2}$ $\rm{s}^{-1}$ (see {\em e.g.} \citep{2008MmSAI..79...19F}). Because $\rm{F_X}\sim \rm{F_{VHE}}$ in gamma-ray binaries, {\em Simbol-X} is able to observe 3 times deeper in the Galactic Plane than HESS. A system with a typical luminosity $\rm{L_X}\sim 10^{33}$ erg/s would be detectable up to the Galactic Center. In spite of it small field of view ($\sim$12'), {\em Simbol-X} would be able to discover serendipitously a couple of gamma-ray binaries. If many more systems are detected, then assuming a power-law distribution N(L>$\rm{L_0}$)$\propto\rm{L_0}^{-\alpha}$ for the luminosity function, this constrains $\alpha$ to exceed 1.\\

In this respect, CTA provides better prospects for serendipitous detections: with 10 times improvement in sensitivity (like {\em Simbol-X}) but with the ability to cover the whole Galatic Plane (like HESS), CTA should detect a dozen new gamma-ray binaries.

\section{Conclusion}

Observations of gamma-ray binaries in hard X-rays will benefit from the unprecedented sensitivity and performances of {\em Simbol-X}. Its higher angular resolution and sensitivity is of major importance for the detection and the identification of new systems in the Galaxy. Precise temporal and spectral measurements in hard X-rays in connection with higher energies will better constrain the existing models. {\em Simbol-X} will play an active role at a time when the {\em Fermi Space Telescope} and CTA will be operating, producing fruitful synergies. Undoubtedly, the discovery of new systems and the understanding of the physics at work in gamma-ray binaries will benefit from {\em Simbol-X}.

%%%%%%%%%%%%%%%%%%%%%%%%%%%%%%%%%%%%%%%%%%%%%%%%
%% BACKMATTER
%%%%%%%%%%%%%%%%%%%%%%%%%%%%%%%%%%%%%%%%%%%%%%%%

\begin{theacknowledgments}
The authors acknowledge support from the {\em European Community} via contract ERC-StG-200911.
\end{theacknowledgments}

\bibliographystyle{aipproc}   % if natbib is available

%\bibliography{b_cerutti}

\end{document}